# Effect of Acoustic Scene Complexity and Visual Scene Representation on Auditory Perception in Virtual Audio-Visual Environments

Stefan Fichna
*Medizinische Physik*
*Universität Oldenburg*
Oldenburg, Germany
stefan.fichna@uol.de

Thomas Biberger
*Medizinische Physik*
*Universität Oldenburg*
Oldenburg, Germany
thomas.biberger@uol.de

Bernhard U. Seeber
*Audio Information Processing*
*Technical University of Munich*
Munich, Germany
seeber@tum.de

Stephan D. Ewert
*Medizinische Physik*
*Universität Oldenburg*
Oldenburg, Germany
stephan.ewert@uol.de

*Abstract*— In daily life, social interaction and acoustic communication often take place in complex acoustic environments (CAE) with a variety of interfering sounds and reverberation. For hearing research and the evaluation of hearing systems, simulated CAEs using virtual reality techniques have gained interest in the context of ecological validity. In the current study, the effect of scene complexity and visual representation of the scene on psychoacoustic measures like sound source location, distance perception, loudness, speech intelligibility, and listening effort in a virtual audio-visual environment was investigated. A 3-dimensional, 86-channel loudspeaker array was used to render the sound field in combination with or without a head-mounted display (HMD) to create an immersive stereoscopic visual representation of the scene. The scene consisted of a ring of eight (virtual) loudspeakers which played a target speech stimulus and non-sense speech interferers in several spatial conditions. Either an anechoic (snowy outdoor scenery) or echoic environment (loft apartment) with a reverberation time (T60) of about 1.5 s was simulated. In addition to varying the number of interferers, scene complexity was varied by assessing the psychoacoustic measures in isolated consecutive measurements or simultaneously. Results showed no significant effect of wearing the HMD on the data. Loudness and distance perception showed significantly different results when they were measured simultaneously instead of consecutively in isolation. The advantage of the suggested setup is that it can be directly transferred to a corresponding real room, enabling a 1:1 comparison and verification of the perception experiments in the real and virtual environment.

*Keywords— complex acoustic environments, psychoacoustics, virtual reality, head-mounted display, virtual acoustic environments*

## I. Introduction

Applications of virtual reality (VR) techniques in hearing research have gained increasing popularity in the recent years due to advancing technologies for both, virtual visual and acoustical representations. VR techniques enable the design and the presentation of ecologically valid complex acoustic environments, which can be used and reproduced under laboratory conditions. In hearing science, ecological validity refers to the degree to which research findings reflect real-life hearing-related function, activity, or participation (see [1], for a review). In this context, virtual representations of CAEs are intended to replicate complex real-life situations to better understand the role of hearing in everyday life. This makes CAEs also important for the evaluation of hearing-supportive devices (e.g., hearing aids or hearables), given that performance gaps between laboratory and real-life situations have been reported for such devices (e.g., [2]).

The classical realistic scenario often referred to in hearing science is the "cocktail party" (e.g., [3]), reflecting normal-hearing (NH) listeners' ability to cope with complex acoustic conditions such as understanding the desired talker while several other interfering talkers are simultaneously speaking. These abilities are often limited in hearing-impaired (HI) listeners (e.g., [4, 5]) and cochlear implant users (e.g., [6]).

Recently, Weisser et al. [7] showed that the complexity of acoustic environments depends mainly on three factors. The first factor, "Multiple acoustic sources distributed in space", depends on the number of acoustic sources, which are representing independent streams of sound information that are competing for the listener's attention. The second factor, "Acoustic source diversity", describes variations in temporal and spectral characteristics but also variations in radiation patterns and position of the sound source (e.g., [8, 9]). The third factor found to influence the complexity of acoustic environments is the "Receiver's task": Participants answered 19 rating questions in complex scenes and for comparison ten more additional questions. The complexity of the scene with additional questions was rated higher and the task was identified as one of the main factors for the complexity of an acoustic scene.

A further factor affecting perception in CAEs is reverberation as typically occurring in enclosed spaces. While reverberation was not dominant enough in [7] to substantially influence scene complexity, other studies found an effect of reverberation on speech intelligibility and listening effort which might lead to a higher complexity compared to anechoic situations. In particular, reverberation reduces temporal modulations of the target speech signal [10] and also reduces the chance to listen into dips of fluctuating interferers. Moreover, the advantage of a spatial separation between the target and masker signals compared to the situation where the target is spatially co-located to the masker is often reduced in the presence of reverberation (e.g., [11, 12, 13]). Besides the previously mentioned detrimental effects of reverberation on speech intelligibility, listening effort also increases with increasing reverberation (e.g., [14]), indicating that more mental effort is required to focus on the desired sound source.

This work was funded by the Deutsche Forschungsgemeinschaft, DFG – Project-ID 352015383 – SFB 1330 C5.

However, for distance perception, listeners seem to take advantage when room information is available [15]. Altogether, reverberation can be considered as a parameter potentially increasing scene complexity.

When using VR techniques to represent CAEs in the laboratory, acoustic rendering can be performed with a loudspeaker array enabling the use of hearing aids [16]. For the representation of the visual scene, head-mounted displays (HMDs) offer a stereoscopic (3-dimensional) view with high contrast ratio. Zhang [17] showed that rendering of scenes with an HMD was rated higher with a screen-based projection system in terms of immersion and the grade of intuitive, interactive and ease of use. However, the presence of the HMD at the head affects the head-related transfer function (HRTF) which can result in a reduction of localization accuracy (e.g., [18]). Gupta et al. [19] showed noticeable changes in the HRTF magnitude spectrum between 1-16 kHz. Also, the interaural level difference (ILD) was significantly different while wearing an HMD. Nevertheless, it remains unclear how changes in the HRTF when wearing an HMD device affect hearing in CAE when listeners behave similar to realistic conditions in which they turn their head, try to understand speech in the presence of interferers and, e.g., estimate the direction of the talker. Moreover, it is unclear in which way the visual representation of the environment displayed by the HMD affects the perception in the CAE.

Starting from the above identified factors, the current study assesses the effect of scene complexity and presence of an HMD on psychoacoustic performance in CAEs: Five psychoacoustic measures were selected which cover four intuitively relevant questions for NH and HI listeners in CAEs. (1) Speech intelligibility addresses the question "What was understood?" (2) Distance and (3) direction perception address the question "Where is the sound source coming from?" (4) Loudness perception ("How loud was the sound source?") is particularly relevant for HI with regard to hearing aid fitting, but also for NH in the context of speech intelligibility as well as for identifying potential threads. (5) Listening effort rates the perceived difficulty of the task and answers to the question "How effortful was it to listen to the sound source?"

Using these five psychoacoustic measures, the current study focuses on three aspects: (i) The effect of wearing an HMD providing a visual scene representation but also affecting HRTFs is assessed. (ii) The effect of the scene factors number of maskers, direction, distance, reverberation and task complexity is tested. (iii) The task complexity is varied by either consecutively testing the five psychoacoustic measures in isolation or by simultaneously assessing all five measures after a single stimulus representation. This additionally shows whether such more time-efficient procedure provides results comparable to those from the slower sequential procedure.

An additional aim of the study was to use a virtual acoustic environment (VAE) that can be compared in a one-to-one fashion to a real-life scenario. For this, a limited number of (virtual) loudspeakers was used as sound sources. In contrast to animated human avatars, which would have to be replaced by voice actors in the real-life environment, the loudspeakers can easily be placed in the virtual as well as in the real scene. The suggested VAE setup may thus help to directly compare hearing in virtual and real environments in future research.

## II. METHODS

### A. Listeners

Twelve normal-hearing listeners aged between 22 and 39 years who were all native German speakers participated in the experiment. Seven of the twelve participants received hourly compensation. The other five listeners were employed by the University of Oldenburg.

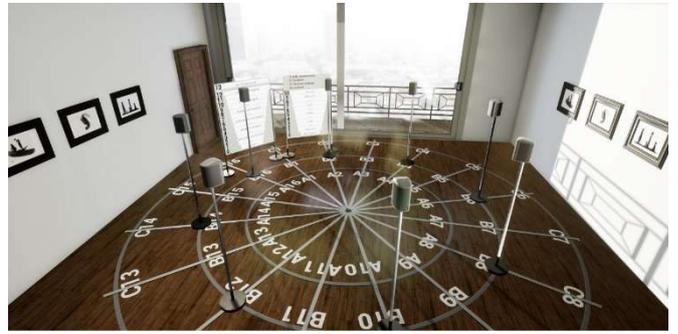

Fig. 1. Visual representation of the reverberant loft apartment/living room in Unreal Engine 4. A circular arrangement of eight loudspeakers was placed around the listener. The net on the floor indicated response alternatives for the perceived distance and direction of the target speaker. Boards represented the response alternatives for the loudness and listening effort ratings.

### B. Audio-visual environments

A reverberant and an anechoic audio-visual environment were created using the room acoustics simulator RAZR ([20], freely available at www.razrengine.com) for the acoustic part and Unreal Engine 4 for the visual part. RAZR calculates early reflections up to the third order using the image source model [21], while late reverberation is calculated by a feedback delay network [22]. An assessment of various common room acoustical parameters and subjective ratings of perceived room acoustical attributes showed a good correspondence between simulated and real rooms [20, 23].

The reverberant environment was a sparsely furnished large loft apartment or living room with a size 12.00 m x 7.50 m x 3.30 m (297 m³) and a resulting reverberation time $T_{60}$ of 1.44 s. The anechoic environment only provided the direct sound. These simulated acoustic rooms were used for both measurements, with HMD and without HMD.

The visual environments were realized in Unreal Engine 4 in accordance with the acoustic properties. The reverberant living room is depicted in Fig. 1. A snowy landscape was designed to fit the anechoic environment. An additional "neutral" visual scene was a virtual version of the anechoic chamber in which the experiments were performed and which was visible for the listeners if no HMD was used. All virtual environments contained a circular array of 8 virtual loudspeakers (separated by 45°) with a radius of 2.4 m centered on the (fixed) position of the listener in the environment. In the virtual reverberant room, the listener was positioned off-center 3.6 m and 3.9 m away from the two closer walls (separated by 7.5 m) and at 2/3 of the longer dimension (12 m), 4 m away from the wall in the front. On the floor, directions (22.5° spacing) and distances (1.7 m, 2.4 m, 3.4 m) from the listener position were indicated by a net (see Fig. 1 and 2) and named by a letter and number combination. Additionally, boards displaying the response alternatives for some of the psychoacoustic measures (loudness and listening effort) were part of the scenes.

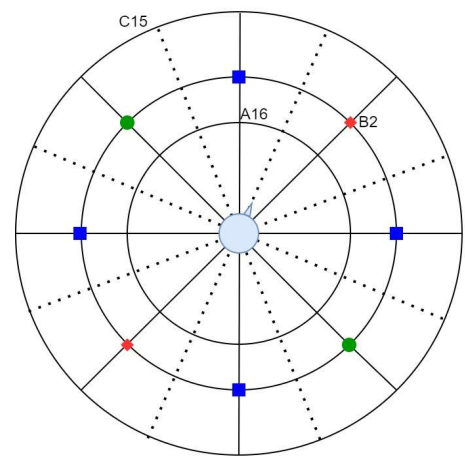

Fig. 2. Illustration of the potential positions of the target signal and maskers. Solid lines: Potential directions of the target signal. Dotted lines: Additional response alternatives offered for the direction perception. The innermost circle A indicates a distance of 1.7 m between the target signal and the listener, while circles B (middle) and C (outmost) represent distances of 2.4 m and 3.4 m, respectively. Green closed circles show the masker positions if 2 maskers were presented. Red diamonds represent the positions of the 2 additional maskers when 4 maskers were presented. Blue squares indicate the positions of the 4 additional maskers when all 8 maskers were active.

*C. Stimuli*

As target signal, sentences from the Oldenburg Sentence Test (OLSA; [24]) were used. The OLSA speech material consists of a large number of meaningless test sentences with correct grammatical structure but no semantical predictability, as all sentences were constructed from a total of 50 words with ten alternatives for each word type (name-verb-numeral-adjective-object). The sentences were spoken by a male talker (mean F0 = 110 Hz) with only a very mild accent. The target signal was presented from direction "16" slightly to the left of the participants' default head orientation (direction "1"), in conditions with known, fixed direction, and was presented from one of the eight potential directions of the target signal for tests with unknown direction (see Fig. 2).

As masker, a male version [25] with the same mean F0 as the target sentences of the International Speech Test Signal (ISTS; [26]) was used. ISTS is nonsense speech generated from six different talkers in different languages. The masker had an overall duration of 54 s, from which randomized tokens of 5 s were extracted for each individual trial.

Complex acoustic environments were created by different spatial arrangements of the target signal and either 2, 4 or 8 uncorrelated, simultaneous ISTS maskers. The target signal was acoustically rendered at one of the 8 different directions of the virtual loudspeakers at 3 possible distances (1.7 m, 2.4 m, 3.4 m). The maskers were always rendered at a distance of 2.4 m (in agreement with the virtual and physical loudspeaker distance) The direction of the maskers always coincided with one of the 8 virtual loudspeakers and depended on the number of active maskers as shown in Fig. 2. The maskers on the positions with the green circles were always active. For four maskers, the maskers on the positions with the red diamonds were added. For 8 maskers, the blue positions were added and all 8 virtual loudspeaker positions were used.

The level of the (anechoic) target signal was 65 dB SPL (A weighted) at a distance of 2.4 m and the level of each individual masker was 62 dB at the distance of 2.4 m. The level of the target signal varied according to distance and was 68 and 62 dB for 1.7 m and 3.4 m, respectively. For the target and masker at the same distance, this resulted in signal-to-noise ratios (SNR) of 0, -3 and -6 dB for 2, 4, and 8 maskers, respectively. RAZR was used to render the stimuli to the physical loudspeaker array (see Sec. E. Apparatus).

*D. Measurement procedure*

Five different psychoacoustic measures were alternatively assessed after the presentation of a target signal in four different tasks:

Speech intelligibility was measured using the Oldenburg Sentence Test [24]. The measurements were conducted as an open sentence test, where the listeners repeated the understood words orally without receiving feedback on the correctness of their response.

Distance and directional perception were combined in one task where the participants identified one of the three potential distances (labeled with A, B, or C, see Fig. 2) and one of the 16 potential directions (labeled 1 to 16) from which the target signal was perceived, e.g., "B2".

The loudness of the target signal was rated using an 11-step-scale based on the Adaptive Categorical Loudness Scaling (ACALOS; [27]). The categories and numbers were indicated on a paper board in the virtual environment when wearing an HMD and in the real environment (VR-Lab) without HMD. The numbers were orally named by the participants. For analysis of the results, the eleven loudness categories were mapped onto the 0-to-50-points categorical units (CU) scale.

The effort to listen to and to understand the target speaker was rated as listening effort using a 14 step-scale based on the Adaptive Categorical Listening Effort Scaling (ACALES; [28]). Like for loudness, the categories and numbers were indicated on a paper board in the virtual and real environment and orally named by the participants.

Additionally, the participants performed the four tasks simultaneously in a separate experimental condition: After a single presentation of a target (and maskers), the participants repeated the sentence of the target speaker, named the position of the target, rated the loudness, and rated the listening effort, always using this order. This task is referred to as SPASE (Speech Perception and Assessment of Spatial Environments) in the following.

The overall five tasks (the four consecutive measurements of speech intelligibility, distance and directional perception, loudness perception, listening effort, as well as SPASE) were performed in 40 partial experiments (see Fig. 3 for an overview). All tasks were performed with four maskers, with and without HMD, in the echoic and anechoic environment. Furthermore, the direction of the target signal was either fixed in a partial experiment block of 10 trials and the participant was advised of the direction (known, fixed position) orally prior to the block as well as visually (when wearing an HMD) by red marks on the grid on the floor, or the direction varied randomly (announced orally prior to the block) between the 8 possible directions (unknown direction). In each of these partial experiment blocks of 10 trials the distance of the target signal was varied. Each direction was presented at least 3 times. In case of partial experiment blocks with unknown target direction, each direction was presented at least once and 2 random directions were presented twice. Additionally, for the SPASE-task, ten trials with 2 maskers and ten trials with 8 maskers were added. This results in overall 480 target sentences in different conditions.

The partial measurements were performed in two measurement sessions on different days and included a familiarization prior to the measurements. A measurement session took about 105 minutes with two blocks of 10 partial experiments separated by a short break. The order of the partial experiments varied per participant, so that of the twelve participants five began the tests first with HMD and seven started without. In addition, six participants started with the echoic and the other six with the anechoic condition, respectively. Likewise, six started with a known direction of the target signal, while the other six participants started with an unknown direction. Each block of 10 partial experiments assessed the psychoacoustic measures in the following fixed order: speech intelligibility, perception of spatial source position, loudness perception and listening effort. Each measurement was first taken either with a known direction or with an unknown direction. After the individual psychoacoustic measures, the simultaneous SPASE measurement was performed.

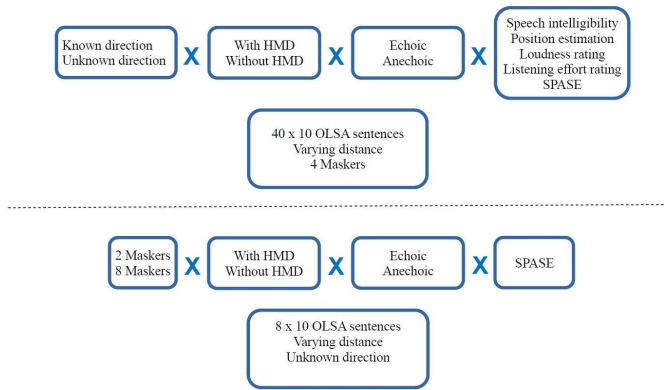

Fig. 3. The top part shows the independent variables for all measurements performed with 4 maskers using the four sequential single tasks as well as SPASE. The bottom part shows the additional measurements with 2 and 8 maskers performed with SPASE with unknown direction of the target.

The familiarization consisted of a total of fifteen trials with varying complexity and lasted about 10 minutes. The familiarization included the simplest scene (anechoic, 2 maskers, frontal target at a distance of 1.7 m) and the potentially most difficult scene (echoic, 8 maskers, target distance 3.4 m from behind) to demonstrate the range for very high and very low listening effort. Three trials comprising 2 maskers (anechoic) with a previously known decreasing target distance (3.4 m, 2.4 m, 1.7 m) were used for distance estimation familiarization. Ten more trials were presented with varying reverberation, masker numbers and different positions of the target signal, so that the listener is familiarized with the range of the loudness and listening effort scales. During the familiarization, the participant was allowed to repeatedly listen to the trial. The position of the target signal could optionally be displayed visually after listening to the stimuli once. The participant received feedback after repeating the sentence of the target speaker in the familiarization.

*E. Apparatus*

The experiments were performed in the VR lab at the University of Oldenburg which is an anechoic chamber including a 3-dimensional array of 86 Genelec 8030 loudspeakers (e.g., [29]). The participants were seated in the center of the loudspeaker array. The main horizontal ring of the array (radius 2.4 m) consists of 48 loudspeakers and is mounted at about head height of the seated participant. Each of the 8 virtual loudspeakers (and sources of the masker sounds) exactly coincides with a physical loudspeaker in the array. Early reflections and late reverberation (simulated using RAZR) were mapped using vector-base amplitude panning (VBAP, [30]) to the 86 loudspeakers, depending on their direction of incidence. The late reverberation used twelve spatially evenly distributed directions (see [20] for further details). For the visualization, the Valve Index was used as stereoscopic head-mounted display [31].

The calibration of the loudspeaker array was carried out by placing a measurement microphone in the center of the array at head height. The delay and broadband level of all individual loudspeakers were adjusted based on sweep measurements (see [32]). Following this calibration, stationary noise with the long-term spectrum of the OLSA speech material was measured from each of the 8 possible target signal directions (loudspeakers in the main ring) in the middle of the array at head height. The level was adjusted to match 65 dB SPL (A weighted) for each target signal direction at 2.4 m distance.

The visual calibration was assessed by a test person wearing the HMD. The test person walked from the center of the loudspeaker array towards one of the loudspeakers in the virtual scene. The point at which the test person's hand with the controller touched the loudspeaker in the virtual world was determined and then the distance between the hand and the real loudspeaker was measured. This procedure was repeated for all 8 virtual speakers. The distance of the virtual loudspeakers and the real loudspeakers differed by ±2 cm.

III. RESULTS

For each of the five psychoacoustic measures speech intelligibility, distance perception, direction perception, loudness perception, and listening effort, a statistical analysis was performed via IBM SPSS using a 5- or 4-way, repeated-measures analysis of variance (ANOVA). Greenhouse-Geisser correction was applied if sphericity was violated. Post-hoc pairwise comparisons used Bonferroni correction. For the analysis and interpretation of the results, the experimental data were grouped in two sets covering a full factorial design for each of the five measures. Statistically significant differences are indicated by asterisks in Fig. 4 and 5. For the loudness ratings, data from two of the twelve listeners were excluded from the analysis and the data plots because large deviations were observed in comparison to all other listeners.

*A. Set 1: Single tasks and SPASE with four maskers*

Fig. 4 shows speech intelligibility, perception of spatial source distance and direction, loudness perception, and listening effort ratings (top to bottom rows) resulting from manipulations (columns) of the factors HMD (with and without HMD), reverberation (echoic and anechoic), task (SPASE and Singletask), distance (target distance of 1.7, 2.4, and 3.4 m), and *a priori* knowledge about the source direction (known versus unknown direction of the target) in the presence of four ISTS maskers (see upper experimental block in Fig. 3). Mean values and inter-individual errorbars at ± one and two standard deviations are shown. For visual guidance, the means are connected by lines. In addition, individual results of the participants are provided as colored closed circles. The asterisks above the connecting lines indicate significant differences (*: $p < 0.5$ **: $p < 0.01$ ***: $p < 0.001$). Speech intelligibility scores are given as percentage correct (word correct rate, top row). Distance ratings are also given as percentage correct (second row). For direction, the azimuth error for the estimated target direction is shown in degree (third row). Loudness ratings are provided in categorical units (CU) where a low value indicates soft and a high CU value indicates a loud target signal (forth row). Listening effort ratings use effort scale categorical units (ESCU), where large values indicate high effort (bottom row).

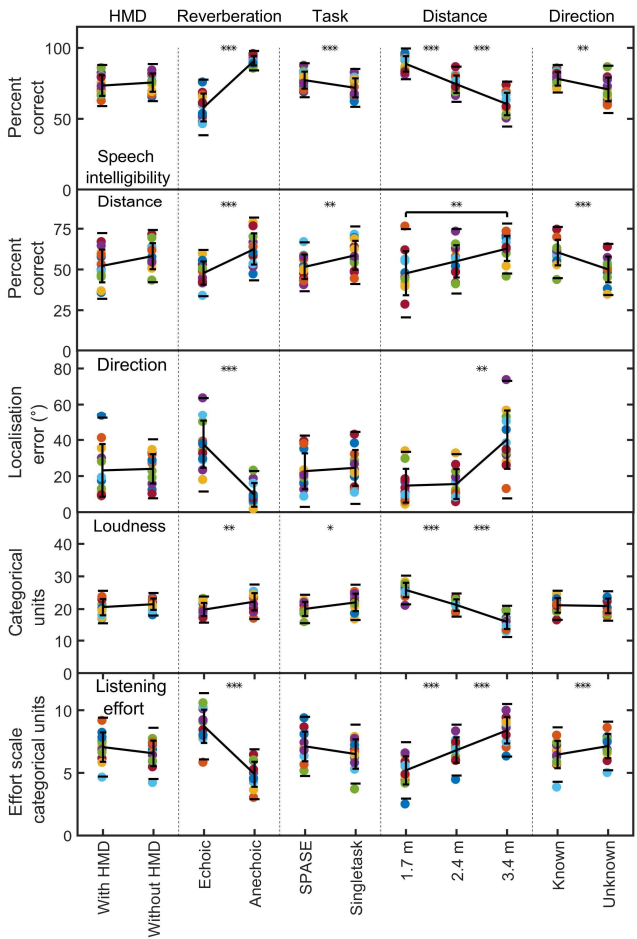

Fig. 4. Mean results and inter-individual standard deviations of the listening test taking into account all measurements in which four maskers were used. The colored closed circles show the average results of each individual participant. Asterisks indicate significant differences *: p< 0.5 **: p< 0.01 ***: p< 0.001.

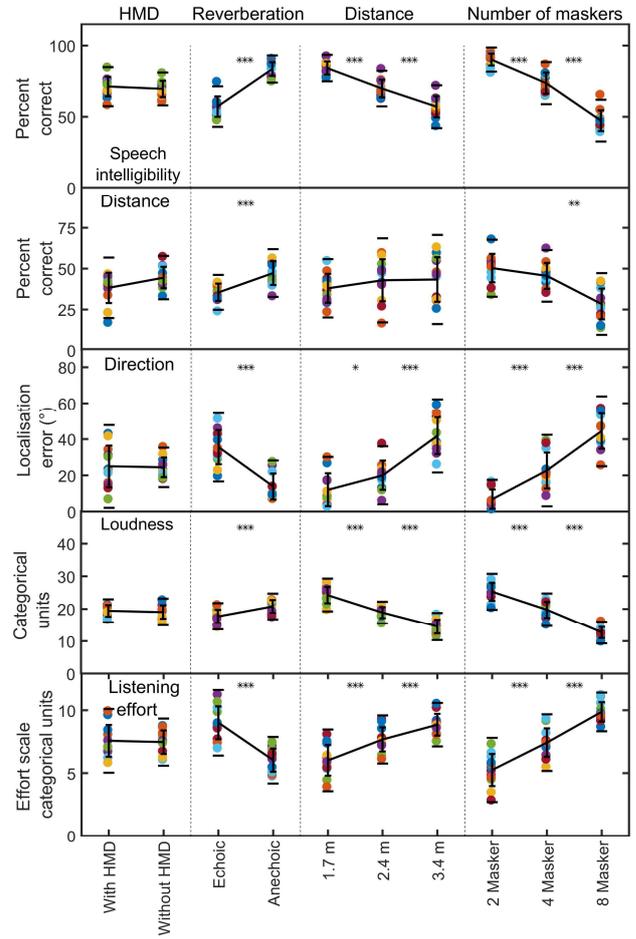

Fig. 5. Mean results and inter-individual standard deviations for all measurements with SPASE and an unknown direction of the target signal. The colored closed circles show the average results of each individual participant. Asterisks indicate significant differences *: p< 0.5 **: p< 0.01 ***: p< 0.001.

In the first column, no significant main effect of HMD was found for any of the five psychoacoustic measures. However, there was a slight tendency for less accurate distance judgements when listeners were wearing an HMD compared to not wearing an HMD with a mean difference (md) of 6 % [$F(1,11) = 3.61$, $p = 0.084$] and for a slightly increased listening effort (higher ratings) when wearing an HMD compared to without HMD [md = 5 %; $F(1,11) = 4.33$, $p = 0.062$]. Also, loudness was estimated slightly lower while wearing the HMD [md = 0.85 CU; $F(1,9) = 4.41$, $p = 0.065$].

The second column shows a significant main effect of reverberation for all five psychoacoustic measures [distance: md = 14.8 %; $F(1,11) = 34.43$, $p < 0.001$; direction: md = 27.92 °; $F(1,11) = 137,30$, $p < 0.001$; loudness: md = 2.4 CU; $F(1,9) = 10.91$, $p = 0.009$]. The effect of reverberation was particularly pronounced for speech intelligibility [md = 33 %; $F(1,11) = 220.81$, $p <0.001$] and for listening effort [md = 3.83 ESCU; $F(1,11) = 132.80$, $p < 0.001$].

For the simultaneous (SPASE) and a sequential (Singletask) measurement (third column), speech intelligibility slightly improved for SPASE [md = 5.38 %; $F(1,11) = 72.15$, $p < 0.001$]. Less accurate distance judgments were observed for SPASE [md = 7.1 %; $F(1,11) = 10.32$, $p = 0.008$] and lower loudness scores were obtained for SPASE [md = 2.03 CU; $F(1,9) = 5.55$, $p = 0.043$]. No significant main effects were observed for direction perception and listening effort.

The results of the fourth column show that the distance had a significant main effect on all psychoacoustic measures. Increasing the distance resulted in a linear decrease of speech intelligibility [md = 14.4 % between 1.7 m and 2.4 m, md = 14 % between 2.4 m and 3.4 m; $F(2,10) = 285.52$, $p < 0.001$], an increase of listening effort [md = 1.62 % and 1.58 %; $F(2,10) = 125.44$, $p < 0.001$], and a decrease of loudness [md = 4.62 CU and 4.95 CU; $F(2,8) = 172.13$, $p < 0.001$]. The accuracy of the target distance perception increased [md = 7.3 % and 7.8 %; $F(2,8) = 10.50$, $p = 0.003$] with increasing distance, while the direction error also increased [md = 0.9 ° and 24.77 °; $F(2,10) = 20.60$, $p < 0.001$] with increasing distance. Post-hoc tests for distance perception showed only a significant difference between 1.7 m and 3.4 m and for direction perception only between 2.4 m and 3.4 m.

The fifth column shows that direction had a significant influence on speech intelligibility [md = 7.5 %; $F(1,11) = 22.63$, $p = 0.001$], distance perception [md = 10.4 %; $F(1,11) = 35.49$, $p < 0.001$] and listening effort [md = 0.69 ESCU; $F(1,9) = 69.19$, $p < 0.001$]. Loudness was not significantly influenced by direction. Direction perception was only measured for unknown target direction, and thus no data are provided in the third row for localization error.

*B. Set 2: SPASE and unknown target direction*

Fig. 5 shows results for the five psychoacoustic measures (rows) in the same manner as in Fig. 4, but based only on the simultaneous SPASE procedure and with unknown target direction. Scene complexity was varied by manipulating the factors HMD (with and without HMD), reverberation (echoic and anechoic), distance (target distance of 1.7, 2.4, and 3.4 m) and additionally the number of maskers (2, 4, and 8), as represented in the four columns (left to right) of Fig. 5.

Results in the first column indicate that there was no significant main effect of HMD for any of the five psychoacoustic measures. There was a trend of slightly more accurate distance perception without HMD than with HMD [md = 6.2 %; $F(1,11) = 3.35$, $p = 0.094$].

As for data set 1 with four maskers, reverberation had a significant main effect on all five psychoacoustic measures (second column), distance [md = 11.8 %; $F(1,11) = 37.70$, $p < 0.001$], direction [md = 21.62°; $F(1,11) = 135.32$, $p < 0.001$], loudness [md = 3.04 CU; $F(1,9) = 27.95$, $p = 0.001$], listening effort [md = 2.98 ESCU; $F(1,11) = 86.85$, $p < 0.001$]. Moreover, a particularly strong effect of reverberation on speech intelligibility [md = 26.4 %; $F(1,11) = 456.34$, $p < 0.001$] was found.

Distance (third column) showed a significant main effect on speech intelligibility [md = 14.42 % between 1.7 m and 2.4 m, md = 12.74 % between 2.4 m and 3.4 m; $F(2,10) = 143.95$, $p < 0.001$], loudness [md = 5.43 CU and md = 4.32 CU; $F(2,8) = 67.65$, $p < 0.001$] and listening effort [md = 1.66 ESCU and md = 1.18 ESCU; $F(2,10) = 116.18$, $p < 0.001$], which were all linearly affected by changes of the target speaker distance. There was also a significant main effect for direction with a detrimental effect mainly visible at the largest distance [md = 8.03 ° and md = 21.81°; $F(2,10) = 136.92$, $p < 0.001$]. Distance had no significant effect on distance perception.

The effect of the number of maskers on the five psychoacoustic measures behaves similarly to the effect of the distance. Here, speech intelligibility [md = 16.56 % between 2 and 4 maskers, md = 26.36 % between 4 and 8 maskers; $F(2,10) = 362.32$, $p <0.001$], loudness [md = 5.55 CU and md = 7.05 CU; $F(2,8) = 62.01$, $p <0.001$], listening effort [md = 2.17 ESCU and md = 2.44 ESCU; $F(2,10) = 128.85$] and direction perception [md = 15.74 ° and md = 21.78°; $F(2,10) = 151.52$, $p <0.001$] were linearly influenced and showed significant main effects. A significant main effect was also found for distance perception [md = 4.7 % and md = 7.2 %; $F(2,10) = 14.56$; $p = 0.001$]. Post-hoc tests revealed, however, that distance estimation was only significantly influenced when the masker number was increased from 4 to 8.

*C. Control condition with neutral visual scene*

An additional control experiment was performed with a sub group of six participants to test the effect of wearing an HMD while displaying the "neutral" virtual representation the VR laboratory in which the measurements took place. Thus, in contrast to the snowy landscape and the living room, the visual scene was uninformative and directly comparable to the visual impression in the real VR laboratory. The results (not shown) were not significantly different to the previous condition with the HMD displaying a congruent scene (landscape and living room) and to the measurements without the HMD. This indicates that the presence of the HMD overall does not affect the psychoacoustic performance, and that performance is independent on the visual information displayed.

## IV. DISCUSSION

In this study, wearing an HMD had no significant effect on the five psychoacoustic measures considered here. This is different to Ahrens et al. [18], where listeners with HMD showed a lower accuracy in estimating lateral sound source locations than listeners without HMD. They explained the increased localization error by HMD-based changes in interaural level and time differences (see Fig. 4 in [18]). Such HMD-based changes in interaural differences are expected to affect sound localization performance in experiments performed under optimal conditions, e.g., using an anechoic room without any masker(s). However, these changes appear to be negligible in complex scenes as used in this study, where the binaural information of the target signal is obscured by interfering masker sources. This is also supported by Kopco et al. [33], where higher localization errors were found for a female speech target in the presence of four spatially distributed male speech maskers than for the target speech without any maskers. As the performance in sound source localization of HI listeners is *per se* often slightly degraded (e.g., [34]), it can be assumed that HMD-based signal distortions are also small for experiments with HI listeners in CAEs. Accordingly, it is assumed that wearing an HMD in CAEs in the here suggested setup, does not significantly influence speech intelligibility, perception of spatial source position, loudness and listening effort ratings of NH and HI listeners. However, the presence of the HMD might alter performance of directional microphones in hearing devices.

In psychoacoustic research, fast and efficient measurement procedures are highly attractive to either reduce measurement time or to increase the amount of data that can be collected within a given time frame. The here suggested SPASE procedure simultaneously assessed five psychoacoustic measures and did not show significant performance differences for listening effort and direction perception compared to the sequential procedure. This seems a bit surprising given that SPASE requires to divide attention to five psychoacoustic measures instead of only one measure. In the current study, the young NH listeners may have primarily focused on understanding the target sentence from which listening effort was derived; further attention to the other psychoacoustic measures did not strongly increase their mental effort. Accordingly, they provided similar effort ratings for SPASE and for Singletask. However, SPASE could have a stronger impact on listening effort ratings in older HI listeners given that some cognitive abilities such as processing speed and memory decline with age (e.g., [35]). Moreover, listeners always performed experiments using the sequential Singletask procedure before they started with SPASE. Accordingly, a training effect might have occurred, compensating for additional mental effort in the SPASE task, and thus resulting in only a small effect of SPASE on listening effort. This training effect is also suspected to be responsible for the (unexpectedly) slightly higher speech intelligibility scores for SPASE in contrast to Singletask. Listeners estimated the target distance more accurately using Singletask instead of SPASE, which may indicate that listeners primarily focused on understanding the target speech and accordingly less attention was given to distance perception. There is no good explanation why loudness of the target speech signal was estimated somewhat but significantly higher for Singletask than for SPASE. However, this difference was only in the region of 2 CU which means that the ratings are in the same loudness category of the CU scale (5 CU steps). Overall, we assume that performance in the SPASE task more accurately reflects real-life performance in the "cocktail party" situation where the listeners perform several tasks at once during a conversation (e.g., "what was said and where was the sound coming from?").

While perceived scene complexity was not substantially influenced by reverberation in [7], reverberation had a significant effect on each of the five perceptual measures considered in the present study, suggesting a contribution to scene complexity. A possible explanation for these different results is that reverberation time varied between 0.2 and 1.2 s in [7], while in this study a reverberation time of 1.44 s was compared to the anechoic scene. Accordingly, the echoic and anechoic scenes in the present study are expected to represent a larger difference that may have resulted in a more prominent effect of reverberation. Another explanation for such differences is that [7] applied a questionnaire to identify factors influencing scene complexity, while in this study the impact of, e.g., reverberation was assessed via psychoacoustic measures. In accordance with other studies (e.g., [11, 12]) listeners had more difficulties in understanding the target speech in echoic than in anechoic room conditions, as reverberation reduced temporal modulation of the target signal and lowered the chance to listen into the dips of the fluctuating masker signals. Moreover, ILDs and ITDs, carrying important information from which listeners benefit in situations where the target speaker is spatially separated from the maskers, are impaired by reverberation, reducing such benefit [12]. The degrading effect of reverberation on binaural cues was also observed in the current results for direction perception, where higher error rates were measured for the echoic than for the anechoic room. Therefore, it can be expected that listeners had more difficulties in localizing and likely in attending to the target speaker in the reverberant scene. Similarly, as observed for direction perception, distance perception was also significantly impaired by reverberation, resulting in less accurate estimations of the perceived distance of the target speaker in the echoic than in anechoic condition. Findings of other studies suspect that the direct-to-reverberant (DRR) energy ratio is an important cue for distance perception in rooms as listeners' distance estimations were more accurate in echoic than in anechoic conditions. One reason why DRR cues may not have played an important role in the current study is because the auditory representation of DRR cues of the target signal was always interfered by at least two maskers, and thus the auditory system had limited access to such cues. From previous studies (e.g., [36, 13]) it is known that reverberation increases listening effort. In agreement with those findings, listeners' ratings indicated more effort for the echoic than for the anechoic environment. Another effect of reverberation concerns the loudness ratings for the target speaker. Here, perceived loudness of the target speaker was increased in the anechoic condition in comparison to the echoic condition. It can be expected that detrimental room reflections (later-arriving room reflections of the target speech plus room reflections of the maskers) and the target speech overlap spectrally and temporally in their auditory representation, effectively reducing time-frequency segments in which the target speech signal is accessible to the auditory system. Accordingly, the reduced loudness of the target speaker in the echoic condition can be assumed to be an effect of masking. Taken together, results of this study showed that reverberation degraded performance of the current young NH listeners. Thus, it can be expected that in the current test setup reverberation contributed to scene complexity. This will likely also be observed in HI listeners.

Several speech intelligibility studies (e.g., [37]) showed that understanding speech becomes more difficult as the number of maskers increases. Moreover, Weisser et al. [7] identified multiple acoustic sources as an important factor for scene complexity. Thus, the effect of the number of maskers on the five psychoacoustic measures was assessed in this study by using 2, 4, and 8 maskers. Given that an increasing number of maskers lowered the SNR, a decrease in speech intelligibility is not surprising. Loudness ratings of the target speaker can be interpreted as an indicator of the target speaker detectability as it is only required to hear and identify the target, but not necessary to understand the target. As shown in the rightmost column of Fig. 5, target loudness was reduced with increasing number of maskers. Thus, increased masking by the fluctuating speech maskers and only fewer spectro-temporal segments of the target speech available to the auditory system resulted in reduced loudness perception. Because of the very similar patterns of results for speech intelligibility and loudness, one could expect that target speaker loudness is a good predictor for speech intelligibility in multiple masker conditions. Recently, Samardzic and Moore [38] successfully predicted speech intelligibility of speech in the steady background noise of a car based on the binaural speech-to-noise loudness ratio. Such metric is assumed to be strongly related to the here measured target speaker loudness in the presence of fluctuating ISTS maskers. The current findings thus motivate further evaluation of loudness models for prediction of speech intelligibility in CAEs. However, a comparison of target speaker loudness and speech intelligibility for anechoic and echoic situations (second column in Figs. 4 and 5) indicates that target speaker loudness seems not sensitive enough as reliable estimator for speech intelligibility in reverberation. As stated in [36], effects of reverberation are negligible for listening in vehicles, and accordingly its influence on the prediction performance of their model was not assessed so far. An increase in the number of maskers also reduced the accuracy for direction and distance estimations. Here, the lower SNRs and increased masking effects cause difficulties to estimate direction and distance of the target. As shown in Fig. 5, listening effort increased with increasing number of maskers and decreasing SNR in line with literature (e.g., [25, 28]) showing that listening effort increases with decreasing SNRs. It should be mentioned that for all factors, represented in the columns of Figs. 4 and 5, a clear inverse relationship between speech intelligibility and listening effort can be observed. Overall, it can be concluded that the number of maskers had a large effect on each of the five psychoacoustic measures in this study. Thus, the number of maskers and decreased SNR appear to be a highly relevant for the complexity of an acoustical scene.

As shown in Fig. 5, changing the target speaker distance had a very similar effect on speech intelligibility, loudness, direction and listening effort as it was observed for variations of the number of maskers. Such a behavior can be expected as variations of target speaker distance and number of maskers effectively change the SNR. Fig. 5 (based on SPASE and unknown target direction) shows that listeners had approximately the same accuracy (about 40 % correct) in estimating the distance of the target speaker for each of the distances of 1.7, 2.4, and 3.4 m. Given that the chance level for correct distance identification was at 33.3 % in this study, an accuracy of 40 % correct indicates that listeners had difficulties in estimating distance. Using four maskers and both known and unknown target directions resulted in significantly more accurate distance estimations for 3.4 m (63 % correct) than for 1.7 m (48 % correct) as observed in Fig. 4. The overall more accurate distance estimations in Fig. 4 than in Fig. 5 might be related to a combined effect of *a priori* knowledge about the target speaker direction improving distance estimations (see rightmost column in Fig. 4) and the presence of eight maskers strongly degrading distance estimation (see rightmost column in Fig. 5).

Although the direction of the interlocutor in many daily-life communication situations, e.g., talking in the office, is known, in crowded conditions with several simultaneously talking and spatially distributed speakers the listener has to switch attention and focus to the desired talker at known and unknown directions. While NH listeners are able to cope with such challenging "cocktail party" conditions, HI listeners often report difficulties in following conversations under such conditions. Therefore, the effect of *a priori* knowledge about the target direction on speech intelligibility, loudness, distance and listening effort was examined in this study. As shown in the last column of Fig. 4, higher speech intelligibility scores were observed when the target speaker came from a known than from an unknown direction. This is not surprising as listeners with *a priori* knowledge can focus on the expected target speaker direction, which makes the task easier compared to the unknown target direction, where listeners have to spot the target speaker direction before attending to it. Consequently, listeners also rated listening effort higher for unknown than for known target speaker directions. The last column of Fig. 4 shows that distance estimations were more accurate for known than for unknown target directions. Here it is expected that for the known direction more processing resources for distance estimation were available than for unknown positions, where cognitive resources are required for both spatial orientation (including head movements) and distance estimation. As expected, loudness perception was not affected by knowledge about the target source direction. Overall, *a priori* knowledge about the target speaker appears to make the listening task easier as attention can be focused on a certain direction and no resources are required for first estimating this direction.

## V. Conclusions

This study suggests an audio-visual test setup with varying complexity of the acoustic environment using virtual reality techniques. By utilizing a relatively low number of 8 loudspeakers as sources, the implemented CAEs can in principle be realized in real rooms as well, allowing for future one-to-one comparisons of listeners' psychoacoustic performance in real and virtual environments.

The effect of (i) wearing an HMD, (ii) several factors affecting scene complexity like number of maskers, unknown direction of the target, distance of the target and reverberation, and (iii) of either consecutively testing five psychoacoustic measures (speech intelligibility, perception of spatial source distance and direction, loudness and listening effort) in isolation or by simultaneously assessing all five measures after a single stimulus representation (SPASE) was assessed. The following conclusions can be drawn:

(1) Wearing an HMD did not affect the evaluation of the examined psychoacoustic measures in the current setup. There was no benefit of the visual scene representation and no disadvantage from altered HRTFs. The results suggest that studies with and without HMD to provide visual cues can be performed in the suggested setup without negatively affecting the basic psychoacoustic performance.

(2) Reverberation had a significant effect on the five psychoacoustic measures, resulting in lower speech intelligibility scores, less accurate distance and direction estimations, lower loudness ratings, and higher listening effort ratings compared to the anechoic room. Reverberation is thus assumed to contribute to scene complexity.

(3) Distance of the target speaker and number of maskers effectively change the SNR between target and maskers and strongly affected the complexity of the acoustical environments. *A priori* knowledge about the target direction increased listeners' performance in speech intelligibility and distance perception, and reduced listening effort compared to conditions with unknown target direction.

(4) The suggested SPASE task, where all five psychoacoustic measures were assessed simultaneously, is suited to reflect the ecologically relevant performance in a complex acoustic environment and to collect data in a time-efficient manner. Effects on perceived loudness, distance and speech intelligibility were observed, however, deviations were generally small and might have been related to training effects.

## Acknowledgment

The authors thank the members of Medizinische Physik and Birger Kollmeier for continued support. The authors also thank Stephan Töpken for remarks on loudness perception. We thank Valve Corporation for providing Valve Index HMDs.